\documentclass[12pt,a4wide]{article}
\usepackage{amsmath,amssymb,amsfonts}

\newcommand{\eqa}{\begin{eqnarray}}
\newcommand{\ena}{\end{eqnarray}}
\begin{document}
\begin{center}
{\large {\bf An Explanation of the ``Pioneer Effect'' based on Quasi-Metric
Relativity}}
\end{center}
\begin{center}
by
\end{center}
\begin{center}
Dag {\O}stvang \\
{\em Institutt for Fysikk, Norges teknisk-naturvitenskapelige universitet, 
NTNU \\
N-7491 Trondheim, Norway}
\end{center}
\begin{abstract}
According to the socalled ``quasi-metric'' framework developed 
elsewhere, the cosmic expansion applies directly to gravitationally
bound systems. This prediction has a number of observable consequences,
none of which are in conflict with observation. In this paper we compare
test particle motion in the nonstatic gravitational field outside a
spherically symmetric source (as predicted by a quasi-metric theory of gravity)
to test particle motion in the Schwarzschild geometry. It is found that if
one incorrectly uses the Schwarzschild geometry (to the relevant accuracy)
to represent the nonstatic quasi-metric model, the largest errors result from 
the mismodelling of null paths. One consequence of this is that using 
electromagnetic signals to track the motion of a non-relativistic particle 
results in the illusion that the particle is influenced by an anomalous 
acceleration of size $cH$ (where $H$ is the Hubble parameter) directed towards
the observer. This result naturally explains the apparently anomalous force 
acting on the Pioneer 10/11, Galileo and Ulysses spacecraft as inferred from 
radiometric data. \\
\end{abstract}
\topmargin 0pt
\oddsidemargin 5mm
\renewcommand{\thefootnote}{\fnsymbol{footnote}}
\section{Introduction}
Some time ago a detailed analysis of the observed versus the calculated orbits
of the Pioneer 10, Pioneer 11, Ulysses and Galileo spacecraft was published 
[1]. The main result was that the observed radiometric data did not agree with
calculations based on standard theory; rather the data indicated the existence 
of an ``anomalous", constant acceleration towards the Sun.

A short summary of the results presented in [1] is as follows:
For the Pioneer spacecraft the Doppler frequency shift of the radio carrier
wave was recorded and analysed to determine the spacecraft's orbits. Two
independent analyses of the raw data were performed. Both showed an anomalous
acceleration towards the Sun, of respectively $(8.09{\pm}0.20){\times}10^{-8}$
cm/s$^2$ and $(8.65{\pm}0.03){\times}10^{-8}$ cm/s$^2$ for Pioneer 10. 
For Pioneer 11 only one result is given; an anomalous acceleration of 
$(8.56{\pm}0.15){\times}10^{-8}$ cm/s$^2$ towards the Sun. The acceleration did
not vary between $40-60$ astronomical units, within a sensitivity of
$2{\times}10^{-8}$ cm/s$^2$.

For the Galileo and Ulysses spacecraft one also got ranging data in addition
to the Doppler data. For Ulysses one had to model the solar radiation pressure
in addition to any constant anomalous acceleration. By doing this it was 
found that Ulysses was influenced by an anomalous acceleration of
$(12{\pm}3){\times}10^{-8}$ cm/s$^2$ towards the Sun, consistent with both
Doppler and ranging data. For Galileo the corresponding result was an
anomalous acceleration of $(8{\pm}3){\times}10^{-8}$ cm/s$^2$ towards the Sun.

Recently a new comprehensive study of the anomalous acceleration was
published [2], including a total error budget for the Pioneer 10 data 
analysis. The new result reported in [2] is an ``experimental'' anomalous 
acceleration of $(7.84{\pm}0.01){\times}10^{-8}$ cm/s$^2$ towards the Sun, and
including bias and uncertainty terms the final value becomes
$(8.74{\pm}1.33){\times}10^{-8}$ cm/s$^2$. For Pioneer 11 an experimental
value of $(8.55{\pm}0.02){\times}10^{-8}$ cm/s$^2$ was given. Also other new 
results, such as annular and diurnal variations in the anomalous acceleration,
were reported in [2].

An interpretation of these results according to the standard general 
relativistic model indicates the existence of an anomalous, time-independent 
force acting on the spacecraft. However there are problems with this 
interpretation since according to the planetary ephemeris there is no 
indication that such a force acts on the orbits of the planets; the 
hypothetical force thus cannot be of gravitational origin without violating 
the weak principle of equivalence. Thus it is speculated that the effect is 
due to anisotropic radiation of waste heat from the radioactive thermal 
generators aboard the spacecraft; the design of the spacecraft is such that 
waste heat may possibly be scattered off the back of the high gain antennae in
directions preferentially away from the Sun [3]. Moreover, besides possible 
anisotropic scattering, an estimate shows that the specific arrangement of 
waste heat radiators on the surface of the spacecraft may perhaps cause 
sufficient anisotropy in the radiative cooling to explain the data [4].
However, it seems that these explanations have been effectively refuted 
[2], [5], [6], [7]. Other possible explanations, such as gas leaks, have been 
proposed [2], [5], but so far it seems that no satisfactory explanation based
on well-known physics exists.

However, it is an intriguing fact that the size of the anomalous acceleration
is of the order $cH$ for all the spacecraft, where $H$ is the Hubble 
parameter. Since this seems to be too much of a coincidence one may suspect 
that the data indicate the existence of new physics rather than a prosaic 
explanation based on standard theory. This has been duly noted by others,
see [2] and references listed therein for a number of new physics suggestions
motivated by the anomaly. But to be acceptable, any non-standard 
explanation should follow naturally from a general theoretical framework. 
In this paper we show that such an explanation can be found, thus the data 
may indeed be taken as evidence for new physics.
\section{A short description of the quasi-metric model}
In references [8], [9] we defined the socalled ``quasi-metric'' space-time 
framework (QMF); this framework is non-metric since it is not based on 
semi-Riemannian geometry. Briefly the geometrical basis of the QMF consists
of a 5-dimensional differentiable manifold with topology ${\cal M}{\times}
{\bf R}_1$, where ${\cal M}={\cal S}{\times}{\bf R}_2$ is a Lorentzian
space-time manifold, ${\bf R}_1$ and ${\bf R}_2$ both denote the real line
and ${\cal S}$ is a compact 3-dimensional manifold (without boundaries). 
Moreover the manifold ${\cal M}{\times}{\bf R}_1$ is equipped with a 
degenerate metric. That is, in addition to the usual time dimension and 3 
space dimensions there is an extra degenerate time dimension represented by 
{\em the global time function} $t$. The physical role of the degenerate 
dimension is to describe global scale changes between gravitational and 
non-gravitational systems. In particular this yields an alternative 
description of the expansion of the Universe. 

The global time function is unique in the sense that it splits quasi-metric 
space-time into a unique set of 3-dimensional spatial hypersurfaces called 
{\em fundamental hypersurfaces (FHSs).} Observers always moving orthogonal to 
the FHSs are called {\em fundamental observers (FOs)}. The topology of
${\cal M}$ indicates that there also exists a unique ``preferred'' ordinary 
time coordinate $x^0$. We use this fact to construct the 4-dimensional 
quasi-metric space-time manifold $\cal N$ by slicing the submanifold 
determined by the equation $x^0=ct$ out of the 5-dimensional differentiable 
manifold. Thus the 5-dimensional degenerate metric field ${\bf g}_t$ may be 
regarded as a one-parameter family of Lorentzian 4-metrics on $\cal N$. Note 
that there exists a set of particular coordinate systems especially well 
adapted to the geometrical structure of quasi-metric space-time, {\em the 
global time coordinate systems (GTCSs)}. A coordinate system is a GTCS iff the
time coordinate $x^0$ is related to $t$ via $x^0=ct$ in ${\cal N}$. Besides, 
for idealized situations it may be possible to find a {\em comoving coordinate 
system}, which by  definition is a (non-static) GTCS where the FOs are at rest.

In reference [10] we introduced a model of the gravitational field outside a 
spherically symmetric, isolated source as predicted by a particular 
quasi-metric theory of gravity developed in [8], [9] (where detailed 
descriptions of this theory can be found). According to this theory
it was found in [10] that at scales similar to the size of the solar system, 
such a gravitational field can be adequately expressed by the 
one-parameter family ${\bf g}_t$ of Lorentzian 4-metrics (expressed in
a spherical comoving coordinate system)
\eqa
ds_t^2=-B(r)(dx^0)^2+({\frac{t}{t_0}})^2{\Big (}A(r)dr^2
+r^2d{\Omega}^2{\Big )},
\ena
where $r$ is a comoving radial coordinate and 
$d{\Omega}^2{\equiv}d{\theta}^2+{\sin}^2{\theta}d{\phi}^2$ is the squared 
solid angle line element. Furthermore $t_0$ represents some arbitrary 
reference epoch setting the scale of the spatial coordinates. (To accurately 
model the solar system gravitational field in a cosmological context one must 
in principle identify the cosmic rest frame with a suitable GTCS and include 
the gravitational effects of the Galaxy and the cosmological substratum.
However, the solar system is so small that we can treat it as an isolated
system moving with constant velocity with respect to the cosmic rest frame.
We can then transform to a new GTCS where the solar system is at rest.)

The locally measured Hubble parameter $H$ is defined as the fractional change 
of $t$ as measured by the FOs [10]. That is, from equation (1) we get
\eqa
H(r,t)={\frac{ct_0}{t}}{\Big(}{\sqrt{B(r)}}{\Big)}^{-1}
{\frac{d}{dx^0}}({\frac{t}{t_0}})={\Big(}{\sqrt{B(r)}}t{\Big)}^{-1}=
{\frac{1}{t}}{\Big (}1+{\frac{r_{\text s0}}{2r}}+
O({\frac{r_{\text s0}^2}{r^2}}){\Big )},
\ena
where $r_{\text s0}$ is the Schwarzschild radius at the arbitrary epoch 
$t_0$, and where we have used equation (6) below in the last step. However, 
for weak gravitational fields the $r$-dependent part of $H$ can be neglected. 
Hence, since we here consider applications of quasi-metric theory to the solar
system only, we will use the approximation $H={\frac{1}{t}}$ for the rest of 
this paper.

General equations of motion are obtained from the geodesic equation using the 
non-metric connection [9]. But these equations of motion {\em cannot} be 
obtained from the geodesic equation using any metric connection. Moreover, as 
shown in [10] special equations of motion for inertial test particles moving 
in the particular metric family (1) take the form (due to the spherical 
symmetry we can restrict the motion to the equatorial plane ${\theta}={\pi}/2$)
\eqa
({\frac{t}{t_0}})^2{\frac{A(r)}{B^2(r)}}({\frac{dr}{cdt}})^2-
{\frac{1}{B(r)}}+{\frac{J^2}{r^2}}=-E,
\ena
\eqa
{\frac{t}{t_0}}r^2{\frac{d{\phi}}{cdt}}=B(r)J,
\ena
\eqa
d{\tau}_t^2=-c^{-2}ds_t^2=EB^2(r)dt^2,
\ena
where $J$ and $E$ are constants of the motion. (Note that the dynamically
measured mass of the central object as measured by distant orbiters increases
to exactly balance the effect on circle orbit velocities of expanding circle 
radii, according to equations (3), (4) and (5). For a further explanation of 
this, see [10].)
By setting the scale factor ${\frac{t}{t_0}}=1$ in equations (1), 
(3), (4) and (5) we recover the equations of motion for inertial test 
particles moving in a spherically symmetric, static gravitational field as 
obtained from General Relativity (GR). Note that $E=0$ for photons and $E>0$ 
for material particles, which may readily be seen from equation (5). 

The functions $A(r)$ and $B(r)$ may be found as series expansions by solving
the field equations, this is done approximately in [10]. For our purposes
we include terms to post-Newtonian order but not higher. Then we have
\eqa
A(r)=1+{\frac{r_{\text s0}}{r}}+O({\frac{r_{\text s0}^2}{r^2}}), {\nonumber} \\
B(r)=1-{\frac{r_{\text s0}}{r}}+O({\frac{r_{\text s0}^3}{r^3}}).
\ena
Note that $A(r)$ and $B(r)$ are not inverse functions [10].
\section{Comparing the quasi-metric and standard models}
Quasi-metric theory predicts that the global cosmic expansion applies directly
to gravitationally bound systems [10]. This has a number of observable
consequences, some of which we will calculate in the following.

We now explore some of the differences between the non-static system described
by equation (1) and the corresponding static system obtained by setting 
${\frac{t}{t_0}}=1$ in (1) and using GR. To begin with we notice that the 
{\em shapes} of free fall orbits (expressed e.g. as functions of the type 
$r({\phi})$) are identical for the two cases [10]. Moreover, it can be shown 
that the time dependence present in equations (3), (4) and (5) does not lead 
to easily observable perturbations in the paths of non-relativistic particles 
compared to the static case [10]. However, as we now illustrate, if one 
considers null paths potential observable consequences appear if one treats 
$r$ as a static coordinate rather than as a comoving one. To simplify matters 
we consider purely radial motion, i.e. $J=0$ 
(one may easily generalize to $J{\neq}0$). Since $E=0$ for photons we get from 
equation (3) that radial null curves are described by the equation
\eqa
{\frac{dr}{dt}}={\pm}{\frac{ct_0}{t}}{\sqrt{\frac{B(r)}{A(r)}}}
={\pm}{\frac{ct_0}{t}}{\sqrt{1-{\frac{2r_{\text s0}}{r}}+
O({\frac{r^2_{\text s0}}{r^2}}}})
={\pm}{\frac{ct_0}{t}}{\Big(}1-{\frac{r_{\text s0}}{r}}+
O({\frac{r^2_{\text s0}}{r^2}}){\Big)},
\ena
the choice of sign depending on whether the motion is outwards or inwards. We
may now integrate (7) along a null path (for convenience we choose the positive
sign in equation (7)). We get
\eqa
{\int}_{{\!}{\!}{\!}r}^{r+R}{\frac{dr'}{1-{\frac{r_{\text s0}}{r'}}+
O({\frac{r_{\text s0}^2}{{r'}^2}})}}=ct_0{\int}_{{\!}{\!}{\!}t}^{t+T}
{\frac{dt'}{t'}}=ct_0{\ln}(1+{\frac{T}{t}})=ct_0{\frac{T}{t}}
{\Big (}1-{\frac{T}{2t}}+O({\frac{T^2}{t^2}}){\Big )},
\ena
where $T$ is the light time along the null path and where $R$ is the radial 
coordinate distance between the object and the observer. From equation (8) we 
find an extra delay, as compared to standard theory, in the time it takes an 
electromagnetic signal to travel from an object being observed to the 
observer. To lowest order this extra time delay is ${\frac{T^2}{2t}}$, and for
weak gravitational fields we may write the extra delay as 
${\frac{HR^2}{2c^2}}$, where $H$ is the Hubble parameter as given from 
equation (2). Besides this extra time delay, the fact that the scale factor in
equation (1) increases with time implies that our model predicts an extra 
redshift, as compared to standard static models, in the Doppler data obtained 
from any object emitting electromagnetic signals. To lowest order this extra 
redshift corresponds to a ``Hubble'' redshift ${\frac{HR}{c}}$. 

But the velocity at any given time of an observed object cannot 
model-independently be split up into one ``ordinary" piece and one ``Hubble" 
piece. This means that there is no direct way to identify the predicted extra 
redshift in the Doppler data. Similarly, at any given time there is no direct 
way to sort out the predicted extra time delay when determining the distance 
to the object. Rather, to test whether the gravitational field is static or 
not one should do observations over time and compare the observed motion to a 
model. In a model one uses a coordinate system and to calculate coordinate 
motions one needs coordinate accelerations. Accordingly we construct the 
``properly scaled coordinate acceleration" quantity $a_{\text c}$. For 
photons this is
\eqa
a_{\text c}{\equiv}{\frac{t}{t_0}}{
\frac{\sqrt{A(r)}}{B(r)}}{\frac{d^2r}{dt^2}}&=&
{\mp}{\frac{c}{t}}+{\frac{t_0}{t}}{\frac{r_{\text s0}c^2}{r^2}}+
O({\frac{r_{\text s0}^2c^2}{r^3}}) {\nonumber} \\
&&={\mp}cH+{\frac{t_0}{t}}{\frac{r_{\text s0}c^2}{r^2}}
+O({\frac{r_{\text s0}^2c^2}{r^3}}).
\ena
The point with this is to show that by treating the comoving coordinate system 
as a static one and using GR, an ``anomalous" term ${\mp}cH$ will be missed 
when modelling coordinate accelerations of photons. 
We see that the sign of the anomalous term is such that the anomalous 
acceleration is oriented in the opposite direction to that of the motion of 
the photons. This means that to sufficient accuracy, treating the comoving 
coordinate system as a static one is equivalent to introducing a variable
``effective'' velocity of light $c_{\text{eff}}$ equal to
\eqa
c_{\text{eff}}=c(1-{\int}_{{\!}{\!}{\!}t}^{t+T}{\frac{dt'}{t'}})=
c(1-HT+O((HT)^2)).
\ena
The change of $c_{\text{eff}}$ with $T$ then yields an anomalous acceleration
\eqa
a_{\text a}={\frac{d}{dT}}c_{\text{eff}}=-cH+O(cH^2T),
\ena
along the line of sight of any observed object. That is, if the comoving
coordinate system is treated as a static one the coordinate motion of any 
object will be observed to slow down by an extra amount if the light time, or 
equivalently, the distance to the observer increases 
and to speed up by an extra amount if the distance decreases. 
Hence, judging from its coordinate motion it would seem as if the 
object were influenced by an anomalous force directed towards the observer.

By integrating the anomalous acceleration over the total observation time 
(i.e. the duration of the experiment) ${\cal T}{\ll}t$ we get an ``anomalous" 
speed
\eqa
w_{\text a}={\int}_{{\!}{\!}{\!}t}^{t+{\cal T}}a_{\text a}dt'=-cH{\cal T}+
O(cH^2{\cal T}^2),
\ena
towards the observer compared to a model where the coordinates are static
rather than comoving. That is, the coordinate motion of any object observed 
over time indicates an anomalous blueshift compared to a model where the 
gravitational field is static. {\em Such an anomalous blueshift may be 
interpreted as an artefact resulting from a mismodelling of the gravitational 
field and the mismodelling of null paths in particular.}

Now, since any observer is typically located at the Earth, the quasi-metric
model in fact predicts that an anomalous acceleration directed towards the 
Earth, rather than towards the Sun, should be seen if one incorrectly 
uses the static GR model to represent the nonstatic gravitational field.
But any directional differences will almost average out over 
time if the observed object moves approximately radially and is located well 
beyond the Earth's orbit. However, compared to a static model where the 
anomalous acceleration is inserted by hand and acts towards the Sun; even if 
directional differences nearly average out there remains a cumulative 
difference. We will calculate this below. If, on the other hand, the line of 
sight to the observed object (e.g. a planet) deviates significantly from the 
radial direction, observations should not be consistent with an anomalous 
acceleration directed towards the Sun. Rather the direction of the anomalous 
acceleration expressed in Sun-centered coordinates would appear to be a 
complicated function of time.

To estimate the predicted differences between a model where the anomalous 
acceleration is towards the Sun and the result given by equation (11);
for the case where the observed object moves approximately radially in the
ecliptic plane and is located well beyond the Earth's orbit it is convenient 
to define the average anomalous acceleration 
${\langle}a_{\text c}{\rangle}_{\text{in}}$ of a photon away from the Sun 
during the time of flight $T$ from the object to the observer. Thus we define
\eqa
{\langle}a_{\text c}{\rangle}_{\text{in}}{\equiv}{\frac{1}{T}}
{\int}_{{\!}{\!}{\!}t}^{t+T}
a_{\text c}dt'={\frac{c}{T}}{\Big [}{\int}_{{\!}{\!}{\!}t}^{t+T_1}-
{\int}_{{\!}{\!}{\!}t+T_1}^{t+T}{\Big ]}{\frac{dt'}{t'}}=
cH{\Big (}2{\frac{T_1}{T}}-1+O(HT){\Big )},
\ena
where $T_1$ is the moment when the photon crosses a plane through the center
of the Sun normal to a line connecting the object and the Sun. (If the Earth
is at the same side of this plane as the object, $T_1=T$.) Similarly we can 
define the average ${\langle}a_{\text c}{\rangle}_{\text{out}}=
-{\langle}a_{\text c}{\rangle}_{\text{in}}$ of a photon towards the Sun during
the time of flight from the observer to the object. We are now able to estimate
the predicted difference ${\delta}a$ between a model where the anomalous 
acceleration is towards the Earth and one where it is towards the Sun. We get
\eqa
{\delta}a=-cH-{\langle}a_{\text c}{\rangle}_{\text{out}}=
-2cH(1-{\frac{T_1}{T}}+O(HT)).
\ena
This function has a minimum at solar conjunction and vanishes when the observed
object and the Earth are at the same side of the Sun. One may show that
${\delta}a$ can be written approximately as a truncated sine function where 
all positive values are replaced by zero. The period is equal to one year and
the amplitude is approximately $2cH{\frac{R_{\text e}}{R_{\text o}}}$, where 
$R_{\text o}$ and $R_{\text e}$
are the radial coordinates of the object and of the Earth, respectively.

Anderson et al. [2] (see also [5]) found an annual perturbation on top of 
the anomalous acceleration $a_{\text P}$ of Pioneer 10. (They claim to see 
such a perturbation for Pioneer 11 also.) Interestingly, the perturbation 
consistently showed minima near solar conjunction [2] (i.e. the 
absolute value of the anomalous acceleration reached maximum near solar 
conjunction). They fitted an annual sine wave to the velocity residuals coming
from the annular perturbation term, using data from Pioneer 10 when the 
spacecraft was about 60 AU from the Sun. Taking the derivative with respect to
time they then found the amplitude of the corresponding acceleration; it was 
found to be $a_{\text{a.t.}}=(0.215{\pm}0.022){\times}10^{-8}$ cm/s$^2$. We 
may compare this to the corresponding amplitude in ${\delta}a$ estimated 
above. We find an amplitude of about $0.29{\times}10^{-8}$ cm/s$^2$
(using the value $8.74{\times}10^{-8}$ cm/s$^2$ for $cH$), close 
enough to be roughly consistent with the data. Besides the annual term a 
diurnal term was also found in the velocity residuals [2], where the 
corresponding acceleration amplitude $a_{\text{d.t.}}$ is large compared to 
$a_{\text P}$. But over one year the contribution from the diurnal term 
averages out to insignificance (less than $0.03{\times}10^{-8}$ cm/s$^2$ [2]).

To find the trajectories of non-relativistic particles we may set 
$E{\equiv}1-{\frac{w^2}{c^2}}$, where ${\frac{w^2}{c^2}}$ is small. 
Then equation (3) yields
\eqa
{\frac{dr}{dt}}={\pm}{\frac{ct_0}{t}}{\sqrt{{\frac{B(r)}{A(r)}}{\Big (}1+
({\frac{w^2}{c^2}}-1)B(r){\Big )}}}=
{\pm}{\frac{ct_0}{t}}{\sqrt{{\frac{r_{\text s0}}{r}}+{\frac{w^2}{c^2}}
+O({\frac{r_{\text s0}^2}{r^2}})}},
\ena
and the properly scaled coordinate acceleration for non-relativistic particles
is
\eqa
a_{\text c}={\mp}{\frac{c}{t}}{\sqrt{{\frac{r_{\text s0}}{r}}+
{\frac{w^2}{c^2}}+O({\frac{r_{\text s0}^2}{r^2}})}}
-{\frac{t_0}{t}}{\frac{r_{\text s0}c^2}{2r^2}}+
O({\frac{r_{\text s0}^2c^2}{r^3}}).
\ena
We see that for non-relativistic particles the effect on coordinate
accelerations of treating the comoving coordinates as static ones and using 
GR, is a factor ${\sqrt{{\frac{r_{\text s0}}{r}}+{\frac{w^2}{c^2}}}}$ smaller 
than the corresponding effect for photons. This means that the trajectories of 
non-relativistic particles do not depend crucially on the fact that the 
gravitational field is non-static. 

On the other hand the paths of photons depend more significantly on whether 
the gravitational field is static or not and this yields the illusion of an 
anomalous acceleration. That is, if one receives electromagnetic signals from 
some freely falling object located e.g. in the outer parts of the solar 
system, the coordinate acceleration of the object as inferred from the signals
should not agree with the ``real" coordinate acceleration of the object if one
treats the comoving coordinates as static ones. Rather, from equation (11) we 
see that it would seem as if the object were influenced by an attractive 
anomalous acceleration of size $cH$. The relevance of this is apparent when 
modeling the orbits of spacecraft and comparing to data obtained from radio 
signals received from the spacecraft; in particular this applies to the 
analyses performed in [1], [2] and [5]. An extra bonus for the model 
considered in this paper is that it predicts small deviations during the year 
if the data are compared to a model where the anomalous acceleration is 
directed towards the Sun rather than towards the Earth. And as we have
seen, this prediction seems to be consistent with the data.
\section{Cosmic expansion and the PPN-formalism}
Orbit analysis of objects moving in the solar system must be based on some
assumptions of the nature of space-time postulated to hold there. The standard
framework used for this purpose is the parameterized post-Newtonian (PPN)
formalism applicable for most metric theories of gravity. But the standard 
PPN-framework does not contain any terms representing expanding space via a
global scale factor as shown in equation (1) since it is inherently assumed 
that the solar system is decoupled from the cosmic expansion. One may try to 
overcome this by inventing some other sense of ``expanding space'' where
the scale factor varies in space rather than in time. But such a model must
necessarily be different from our quasi-metric model, and we show below that
it cannot work.
Thus, to illustrate the inadequacy of the PPN-framework to model 
expanding space we now consider a specific model where suitable terms are 
added by hand in the metric. We may then compare to the change in light 
time obtained from our quasi-metric model.

One may try a post-Newtonian metric of the type
\eqa
ds^2=-{\Big(}1-{\frac{r_{\text s}}{r}}+{\frac{2H_0r}{c}}+
O({\frac{r_{\text s}^3}{r^3}}){\Big)}(dx^0)^2+
{\Big(}1+{\frac{r_{\text s}}{r}}-{\frac{2H_0r}{c}}+
O({\frac{r_{\text s}^2}{r^2}}){\Big)}dr^2+r^2d{\Omega}^2,
\ena
where $H_0$ is a constant, to describe expanding space within the 
PPN-framework. (To show that this metric yields a spatially variable
scale factor, transform to isotropic coordinates.) It may be readily shown
that the metric (17) yields a constant anomalous acceleration $cH_0$
towards the origin. But the problem with all metrics of this type is that 
they represent a ``real'' anomalous acceleration of gravitational origin, and 
this is observationally excluded from observations of planetary orbits [2].

Anyway we may calculate the change in light time ${\Delta}T$ due to the 
terms containing $H_0$ in equation (17) by integrating a 
radial null path from $r_{\text o}$ to $r_{\text e}$ (let 
$r_{\text o}>r_{\text e}$, say). This yields
\eqa
{\Delta}T=-c^{-2}H_0(r_{\text o}^2-r_{\text e}^2)+{\cdots}{\approx}-H_0T^2,
\ena
where the light time $T$ is equal to $c^{-1}(r_{\text o}-r_{\text e})$ to 
first order and where the last approximation is accurate only if $r_{\text o}$
is large. Note that ${\Delta}T$ is negative; this is quite counterintuitive
for a model representing expanding space. 

Anderson et al. [2] have considered a phenomenological model representing 
``expanding space'' by adding a quadratic in time term to the light time in 
order to determine the coefficient of the quadratic by comparing to data. 
To sufficient accuracy this model may be represented by the transformation
\eqa
T{\rightarrow}(1+a_{\text{quad}}t)T{\equiv}T + {\Delta}T,
\ena
where $a_{\text{quad}}$ is a ``time acceleration'' term. This model
fits both Doppler and range very well [2].

If we compare equations (18) and (19) we see that the value
$a_{\text{quad}}=-H_0{\frac{T}{t}}$ corresponds to the change in light time
calculated from the metric (17). This is far
too small (of order $10^{-30}$ s$^{-1}$ for a light time of a few hours)
to be found directly from the tracking data. Thus, the fact that
$a_{\text{quad}}$ was estimated to be zero based on the tracking data alone 
[2] in no way favours a constant acceleration model over a time acceleration 
model. They are in fact equivalent as far as the data are concerned.

To compare differences in $a_{\text{quad}}$ we may use equation (8) to 
find the extra delay ${\frac{H}{2c^2}}(r_{\text o}-r_{\text e})^2=
{\frac{HT^2}{2}}$ in the light time compared to the static case. This 
corresponds to a value $a_{\text{quad}}={\frac{H^2T}{2}}$
for the time acceleration term. But a determination of $a_{\text{quad}}$ 
directly from the tracking data still reflects the model-dependency explicitly 
present in the orbit determination process. This means that a determination 
of $a_{\text{quad}}$ directly from the tracking data should in principle be 
consistent with the model (17) and not with our quasi-metric model. However,
the quantity $a_{\text{quad}}$ is so small that it is not feasible to check 
this. But the fact that a phenomenological model of the type (19) works so well
should be taken to mean that the explanation of
the anomalous acceleration given in this paper is sufficient.
\section{Conclusion}
We conclude that a natural explanation of the data is that the gravitational 
field of the solar system is not static with respect to the cosmic expansion.
This also explains why any orbit analysis program based on the 
PPN-formalism is insufficient for the task and how the largest errors
arise from the mismodelling of null paths. (In fact, using the PPN-formalism
is equivalent to introducing a variable ``effective'' velocity of light as 
shown in equation (10).) But these explanations, while 
not involving any ad hoc assumptions, are based on the premise that 
space-time is quasi-metric. That is, rather than being described
by one single Lorentzian metric, the gravitational field of the solar system 
should be modeled (to a first approximation) by the metric family shown in 
equation (1). From a theoretical point of view this premise is radical; thus 
it is essential that the subject is further investigated to make certain that 
more mundane explanations may be eliminated. However, so far no such 
explanations based on well-known physics have been found. But the facts are 
that the model presented in this paper follows from first principles and fits 
the data very well. Moreover there exists independent observational 
evidence in favour of the prediction that the Earth-Moon system is not static 
with respect to the cosmic expansion, and quasi-metric gravity predicts these 
observations from first principles as well [10]. Thus the fact is that several
observations in the solar system seem to involve the Hubble parameter. 
This should not be dismissed as a coincidence, and indicates 
that explanations based on quasi-metric relativity should be taken seriously.
\\ [4mm]
{\bf Acknowledgement} \\ [1mm]
I wish to thank Dr. K{\aa}re Olaussen for helpful discussions and thoughtful
comments on the manuscript.
\\ [4mm]
{\bf References} \\ [1mm]
{\bf [1]} J.D. Anderson, P.A. Laing, E.L. Lau, A.S. Liu, M.M. Nieto,
S.G. Turyshev,  \\
{\hspace*{6.3mm}}{\em Phys. Rev. Lett.} {\bf 81}, 2858 (1998)
(gr-qc/9808081). \\
{\bf [2]} J.D. Anderson, P.A. Laing, E.L. Lau, A.S. Liu, M.M. Nieto,
S.G. Turyshev,  \\
{\hspace*{6.5mm}}{\em Phys. Rev. D} {\bf 65}, 082004 (2002)
(gr-qc/0104064). \\
{\bf [3]} J.I. Katz, {\em Phys. Rev. Lett.} {\bf 83} 1892 (1999) 
(gr-qc/9809070). \\
{\bf [4]} E.M. Murphy, {\em Phys. Rev. Lett.} {\bf 83} 1890 (1999)
(gr-qc/9810015). \\
{\bf [5]} S.G. Turyshev, J.D. Anderson, P.A. Laing, E.L. Lau, A.S. Liu, 
M.M. Nieto, \\
{\hspace*{5.1mm}} gr-qc/9903024 (1999). \\
{\bf [6]} J.D. Anderson, P.A. Laing, E.L. Lau, A.S. Liu, M.M. Nieto,
S.G. Turyshev, \\
{\hspace*{6.4mm}}{\em Phys. Rev. Lett.} {\bf 83}, 1893 (1999)
(gr-qc/9906112). \\
{\bf [7]} J.D. Anderson, P.A. Laing, E.L. Lau, A.S. Liu, M.M. Nieto,
S.G. Turyshev, \\
{\hspace*{6.4mm}}{\em Phys. Rev. Lett.} {\bf 83}, 1891 (1999)
(gr-qc/9906113). \\
{\bf [8]} D. {\O}stvang, gr-qc/0112025 (2001). \\
{\bf [9]} D. {\O}stvang, {\em Doctoral thesis}, (2001) (gr-qc/0111110). \\
{\bf [10]} D. {\O}stvang, gr-qc/0201097 (2002). 
\end{document}